# Deep learning-based multi-level airway collapse site localization on obstructive sleep apnea patients based on snoring sounds

Ying-Chieh Hsu[1,2,3], Stanley Yung-Chuan Liu[4], Chao-Jung Huang[5], Chi-Wei Wu[6], Ren-Kai Cheng[6], Jane Yung-Jen Hsu[6,7], Shang-Ran Huang[8], Yuan-Ren Cheng[8], Fu-Shun Hsu[9,10,11]*

1 Department of Biomedical Engineering, National Taiwan University, Taipei, Taiwan
2 Department of Otolaryngology-Head and Neck Surgery, Taipei Tzu Chi Hospital, Buddhist Tzu Chi Medical Foundation, Taipei, Taiwan; iammed91@gmail.com
3 School of Medicine, Tzu Chi University, Hualien, Taiwan
4 Department of Oral & Maxillofacial Surgery, Nova Southeastern University College of Dental Medicine and College of Allopathic Medicine, Florida, USA; sliu@nova.edu
5 All Vista Healthcare Center and Center for AI and Advanced Robotics, National Taiwan University, Taipei, Taiwan; cjhuang0717@gmail.com
6 Department of Computer Science and Information Engineering, National Taiwan University, Taipei, Taiwan; aladar.wu@gmail.com; b06902127@ntu.edu.tw; yjhsu@csie.ntu.edu.tw
7 Department of Artificial Intelligence, Chang Gung University College of Intelligent Computing, Taoyuan, Taiwan
8 Heroic Faith Medical Science Co., Ltd., New Taipei, Taiwan; srh623@gmail.com; infie.cheng@heroic-faith.com
9 School of Medicine, Chung Shan Medical University, Taichung, Taiwan; fshsu@heroic-faith.com
10 Department of Internal Medicine, Chung Shan Medical University, Taichung, Taiwan
11 Hospital Development Affairs Office, Chung Shan Medical University Hospital, Taichung, Taiwan

Short title: Multi-level OSA Acoustic Monitoring

*Correspondence: fshsu@heroic-faith.com; Tel.: 886-2-2659 0291

**Abstract:**

**Objective** This study evaluated deep learning models, specifically ResNet-50 and Audio Spectrogram Transformer (AST), to identify multi-level upper airway collapses in obstructive sleep apnea (OSA) patients using snoring sounds.

**Methods** Snoring recordings from 37 subjects undergoing drug-induced sleep endoscopy (DISE) between 2020 and 2021 were used to train models for two multi-label tasks: identifying obstructions at the Velum (V), Oropharynx (O), Tongue-base (T), and Epiglottis (E) levels, and the retropalatal (RP) vs. retroglossal (RG) levels.

**Results** The AST model significantly outperformed ResNet-50 ($P < 0.05$ after Bonferroni correction), effectively identifying V-, O-, and RP-level obstructions (F1-score: 0.59–0.87, Mathiew's Correlation Coefficient: 0.50–0.80, and AUC: 0.87–0.97). However, the performance of both models at the T, E, and RG levels remained limited due to small sample sizes, even with the Synthetic Minority Over-sampling Technique. A full-night recording analysis further demonstrated the feasibility of acoustic profiling for airway dynamics.

**Conclusion** The AST models can effectively identify airway obstruction at V, O, and RP levels via snoring sounds. The proposed models offer a promising non-invasive tool for tracking overnight airway dynamics and assisting in clinical treatment planning.

**Keywords:** audio spectrogram transformer; deep learning; drug-induced sleep endoscopy; obstructive sleep apnea; VOTE classification

## 1. Introduction

Obstructive sleep apnea (OSA) manifests as recurrent episodes of apnea or hypopnea during sleep[1]. The cause of OSA is repetitive partial or complete soft-tissue collapse at different levels of the upper airway, obstructing breathing airflow during sleep. The obstruction may occur at a single position only or concurrently at multiple levels of the upper airway[2]. The obstruction sites can be roughly dichotomized into retropalatal (RP) and retroglossal (RG)[1], or they can be further divided and categorized on the basis of clinical classification systems, such as the Velum (V), Oropharynx (O), Tongue Base (T), and Epiglottis (E) classification system, collectively referred to as the VOTE system[3].

Identifying the level of upper airway collapse in OSA patients is clinically important because the effectiveness of different OSA treatments is reportedly sensitive to the site of obstruction. For instance, OSA patients with epiglottic collapse often exhibit poor adherence to continuous positive airway pressure (CPAP) therapy[4]. CPAP therapy used in treating tongue base or epiglottis collapse is not as effective as in dealing with oropharyngeal wall obstruction[5]. Oral appliances using a mandibular advancement device are more suitable for individuals with tongue base obstruction, and they may aggravate complete concentric collapse at the palate level and complete lateral oropharyngeal collapse[6]. Hypoglossal nerve stimulation is reported to be ineffective in treating complete concentric collapse at the palatal level[7].

Polysomnography (PSG) is the gold standard for diagnosing OSA, yet it provides little information about the anatomical sites of upper airway blockages[1, 8]. In the clinic, several techniques are used to identify the site of obstruction in patients with OSA. Müller maneuver[9] and drug-induced sleep endoscopy (DISE)[10] use minimally invasive nasofibroscopy to observe the airway collapse. However, they did not reflect actual conditions experienced by OSA patients during natural sleep[11-13]. Non-invasive imaging techniques, such as computed tomography (CT)[14], sleep video-fluoroscopy (SVF)[15], and dynamic magnetic resonance imaging (MRI)[16], are often limited by wake-state discrepancies, radiation exposure, high costs, labor-intensive setup, or a lack of temporal information, making them unsuitable for longitudinal monitoring.

Alternatively, many studies investigate the automatic localization of airway collapse by analyzing the snoring sound. Snoring, the hallmark symptom of OSA, arises from the vibration of collapsed soft tissues during sleep[17]. Snoring sounds generated at different airway levels are reported to have distinct acoustic characteristics[18-20]. Thus, training a machine learning or deep learning model with the acoustic features of snoring sounds offers a promising avenue for pinpointing the precise sites of obstruction[21, 22]. So far, many studies have proposed models for locating single-level obstructions. However, multi-level obstructions can present in up to approximately 65% of OSA patients[23]. Some studies investigate multi-level obstruction localization, but they all rely on machine learning models, such as logistic regression, k-nearest neighbors, and support vector machines (SVMs)[24-26].

Given this gap, the current study aimed to develop deep-learning models capable of localizing multi-level obstructions based on snoring sound clips. We took a multi-label classification[27] approach to finetune ResNet-50[28] and an audio spectrogram transformer (AST)[29] to identify airway obstruction 1) at VOTE levels and 2) in the RP and RG spaces based on the snoring sounds of OSA patients. To address the inherent class imbalance in the VOTE snoring dataset, the synthesized minority oversampling technique (SMOTE)[30] was employed. The models' performance was statistically compared. We further displayed the model's potential for longitudinal tracking of airway collapse dynamics by analyzing a full-night tracheal sound recording.

## 2. Related Work

Earlier studies on the correlation between snoring sounds and upper airway collapse commonly differentiate between palatal and non-palatal snoring[31]. Recent advancements have focused on training a machine-learning or deep-learning model to classify snoring sounds based on VOTE classification ground truth established through a DISE procedure. Many studies have researched this topic using a publicly available snoring sound dataset, the Munich Passau Snore Sound Corpus (MPSSC), introduced at the INTERSPEECH 2017 Computational Paralinguistics Challenge[32]. The MPSSC dataset contains 828 snore samples (V:484, O: 216, T: 39, and E: 89). Previous efforts attained an unweighted average recall (UAR) between 67.0%–87.5% based on SVM, naïve Bayesian models, and deep neural networks[33-39]. Besides, some studies developed models on a private dataset[40-43]. Nevertheless, the above studies excluded the snoring sounds generated at multiple levels. By contrast, few studies classified snoring sounds involving multi-level obstruction [24-26]; however, they did not investigate the four types of VOTE classification. Table 1 summarizes previous related work and its performance and limitations.

**Table 1. Summary and comparison of the literature on snoring sound localization techniques.**

| Year | Author | Model / Architecture | Classification Task | Dataset | Performance | Limitation/Note |
|---|---|---|---|---|---|---|
| 1999 | Osborne et al.[31] | Thresholding; Peak factor ratio | Palatal vs. Non-palatal | Private | SEN: 100%, SPE: 100% | Simple acoustic heuristic; extremely small sample size; no separate test set |
| 2016 | Lee et al.[24] | Logistic Regression | Multi-level or not; Single-level V, O, T, E | Private | Multi-level SEN: 93%; SPE: 67% Single-level SEN: 69%-92%; SPE: 60%-73% | Subject-wise; no separate test set mentioned |
| 2016 | Qian et al.[40] | Random Forest | Single-level V, O, T, E | Private | UAR: 78% | Excluded multi-level samples |
| 2016 | Qian et al.[42] | SVM | Single-level V, O, T, E | Private | UAR: 71.2%; WAR: 78.2% | Excluded multi-level samples |
| 2016 | Schmitt et al.[43] | SVM; BoAW | Single-level V, O, T, E | Private | UAR: 79.5%; WAR: 79.7% | Excluded multi-level samples |
| 2017 | Schuller et al.[32] | SVM | Single-level V, O, T, E | MPSSC | UAR: 58.5% | Excluded multi-level samples |
| 2017 | Amiriparian et al.[34] | AlexNet & VGG19 | Single-level V, O, T, E | MPSSC | UAR: 67.0% | Excluded multi-level samples |
| 2018 | Demir et al.[33] | SVM | Single-level V, O, T, E | MPSSC | UAR: 72.6% | Excluded multi-level samples |
| 2018 | Vesperini et al.[36] | DNN | Single-level V, O, T, E | MPSSC | UAR: 67.7% | Excluded multi-level samples |
| 2018 | Qian et al.[41] | DNN | Single-level V, O, T, E | MPSSC | UAR: 72.8% | Excluded multi-level samples |
| 2019 | Qian et al.[44] | Naïve Bayes; BoAW | Single-level V, O, T, E | MPSSC | UAR: 69.4% | Excluded multi-level samples |
| 2019 | Schmitt et al.[35] | CNN & LSTM | Single-level V, O, T, E | MPSSC | UAR: 67.0% | Excluded multi-level samples |
| 2020 | Sun et al.[38] | SVM; trend-based MFCC | Single-level V, O, T, E | MPSSC | UAR: 87.5% | Excluded multi-level samples |
| 2020 | Zhang et al.[25] | SVM | Multi-level V, O, E, V + T/E, V + O, VOTE, and non-snoring | Private | ACC: 89.8% | Multiclass classification without all possible combinations |
| 2022 | Ding et al.[39] | CNN | Single-level V, O, T, E | MPSSC | UAR: 77.1% | Excluded multi-level samples |
| 2022 | Liu et al.[26] | KNN | RP, RG, RP+RG | Private | ACC: 85.55%; F1-score: 0.85 | Did not use VOTE classification |
| **2026** | **This study** | **ResNet-50 & AST** | **Multi-label V, O, T, E, & RP vs RG** | **Private** | **F1-score: 0.59 (V), 0.76 (O) MCC: 0.50 (V), 0.67 (O) AUC: 0.87 (V), 0.92 (O)** | **First to achieve multi-label detection with AST** |

Note: SVM: support vector machine; BoAW = Bag-of-Audio-Word; DNN: deep neural networks; LSTM: long short-term memory networks; MFCC: Mel-frequency Cepstral Coefficients; CNN: convolutional neural networks; KNN: k-nearest neighbors; AST: audio spectrogram transformer; V: velum, O: oropharyngeal wall, T: tongue base; E: epiglottis; RP: retropalatal; RG: retroglossal; SEN: sensitivity; SPE: specificity; ACC = Accuracy; UAR = Unweighted Average Recall; WAR = Weighted Average Recall; MCC: Mathiew's Correlation Coefficient; AUC: area under the operating characteristic curve.

## 3. Materials and Methods

### *3.1 Subjects*

A cohort comprising 39 individuals aged 20 to 70, diagnosed with OSA, was enrolled in outpatient otolaryngology clinics at Taipei Tzu-Chi Hospital, Taipei, Taiwan, between 2020 and 2021. The inclusion criteria were 1) a diagnosis of OSA, 2) age ≥20 years, 3) ability to provide informed consent, and 4) intolerance to continuous CPAP therapy. Exclusion criteria encompassed: 1) a history of prior OSA surgery, 2) American Society of Anesthesiologists class 3 or above, or 3) vulnerable populations, including pregnant women, prisoners, individuals with cognitive impairment or mental illness affecting capacity to consent, and adults under guardianship. Diagnosis of OSA was corroborated through attended PSG, adhering to the standardized protocols outlined by the American Academy of Sleep Medicine[45]. Before participation, all enrolled subjects provided written informed consent per the guidelines established by the Research Ethics Review Committee of Taipei Tzu-Chi Hospital (permit number: 09-XD-079).

### *3.2 Drug-Induced Sleep Endoscopy and Snoring Sound Recording*

DISE was conducted within an operating theater under the supervision of a seasoned otolaryngologist. Sedation was induced through intravenous administration of propofol utilizing a target-controlled infusion system. The bispectral index was maintained between 65 and 70 throughout the DISE procedure to ensure optimal depth of anesthesia.

During the DISE procedure, the subjects’ breathing sounds were recorded using a digital stethoscope (AccurSound AS-101, Heroic Faith Medical Science Co., Ltd., New Taipei, Taiwan) connected to a smartphone (Mi 9T pro, Xiaomi, Beijing, China) with a custom recording app installed. The acoustic patch of the digital stethoscope was securely affixed to the subjects' submental region with tape. The breathing sounds were sampled at a rate of 4 kHz with a bit depth of 16 bits using an AccurSound AS-101; however, the audio signal was transmitted to a smartphone and saved as a WAVE format (.wav) file in IEEE 754 single-precision (32 bits). The DISE videos and the breathing sounds were manually synchronized.

*3.3 VOTE Labeling on DISE videos*

Periods of DISE videos exhibiting clear visibility were selected for the subsequent VOTE scoring, while those with excessive subject salivation or poor endoscopic views were excluded. These selected DISE periods were independently scored by three experienced otolaryngologists (with sleep surgery fellowship training), and the ground-truth VOTE labels were established based on a majority consensus (at least two out of three). The endoscopic video snapshots on the left side of Fig 1 display partial upper airway obstruction at the Velum (Fig 1a), Oropharynx (Fig 1b), Tongue Base (Fig 1c), and Epiglottis (Fig 1d) levels. The periods involving single-level obstruction were labeled "V," "O," "T," or "E"; the periods involving multi-level obstruction were designated, e.g., "VO," "VT," "VOT," and so forth. These labels were referred to as VOTE labels.

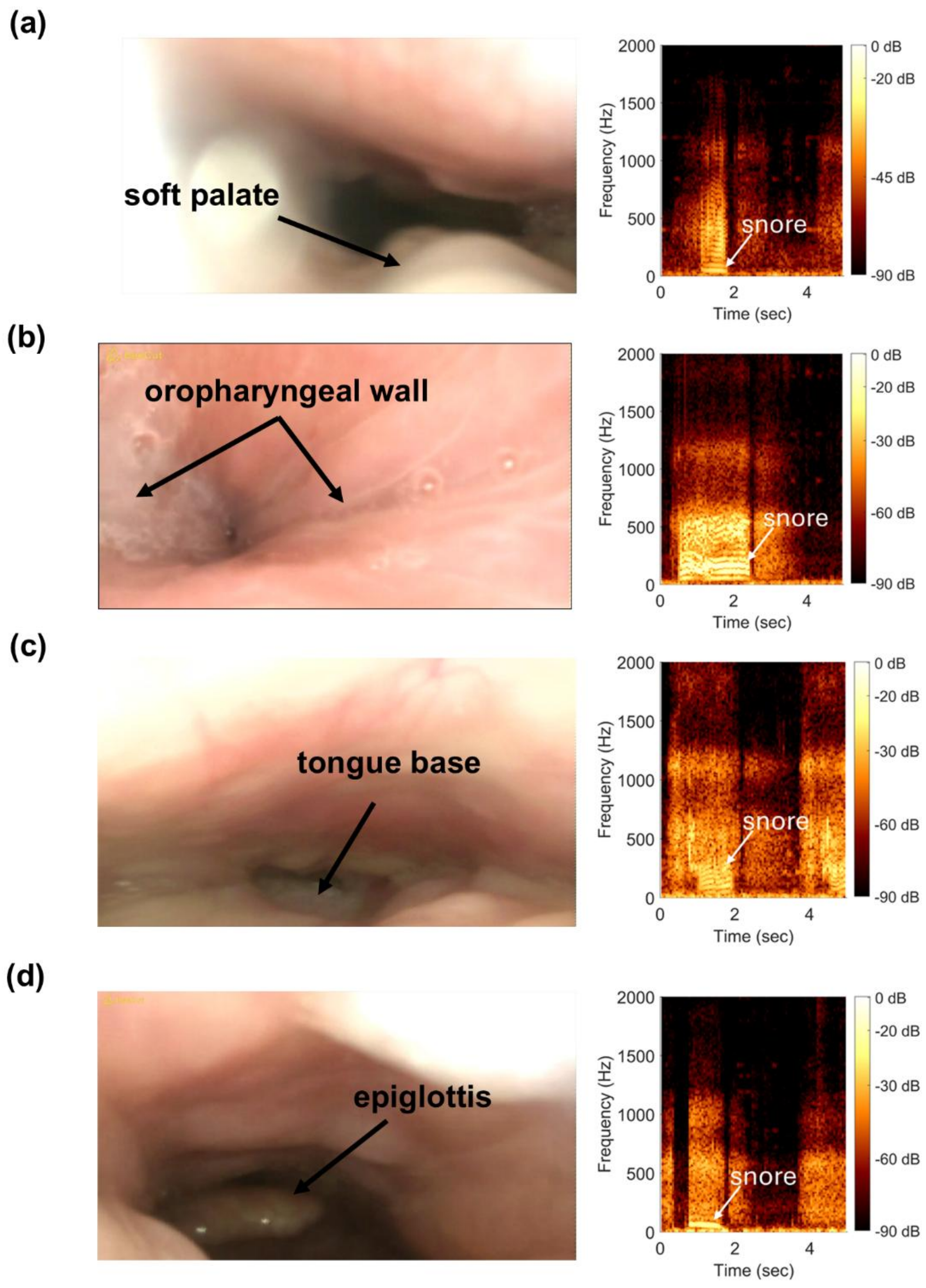

**Figure 1. The classification of obstruction sites and their corresponding spectrogram.** The endoscopic snapshots on the left side show the obstructions classified at the (a) Velum (V), Oropharynx (O), (c) Tongue Base (T), and (d) Epiglottis (E) levels. The graphs on the right display the temporal-spectral representation (spectrogram) of the snoring sounds corresponding to the obstructions. White arrows indicate the features of the snoring sounds.

*3.4 Snore Labeling on Breath Sound Recordings*

The recorded breathing sounds were subsequently transformed into spectrograms (Fig 2a) using custom labeling software[46]. Three board-certified respiratory therapists annotated the start and end times of snoring events (gray periods in Fig 2b).

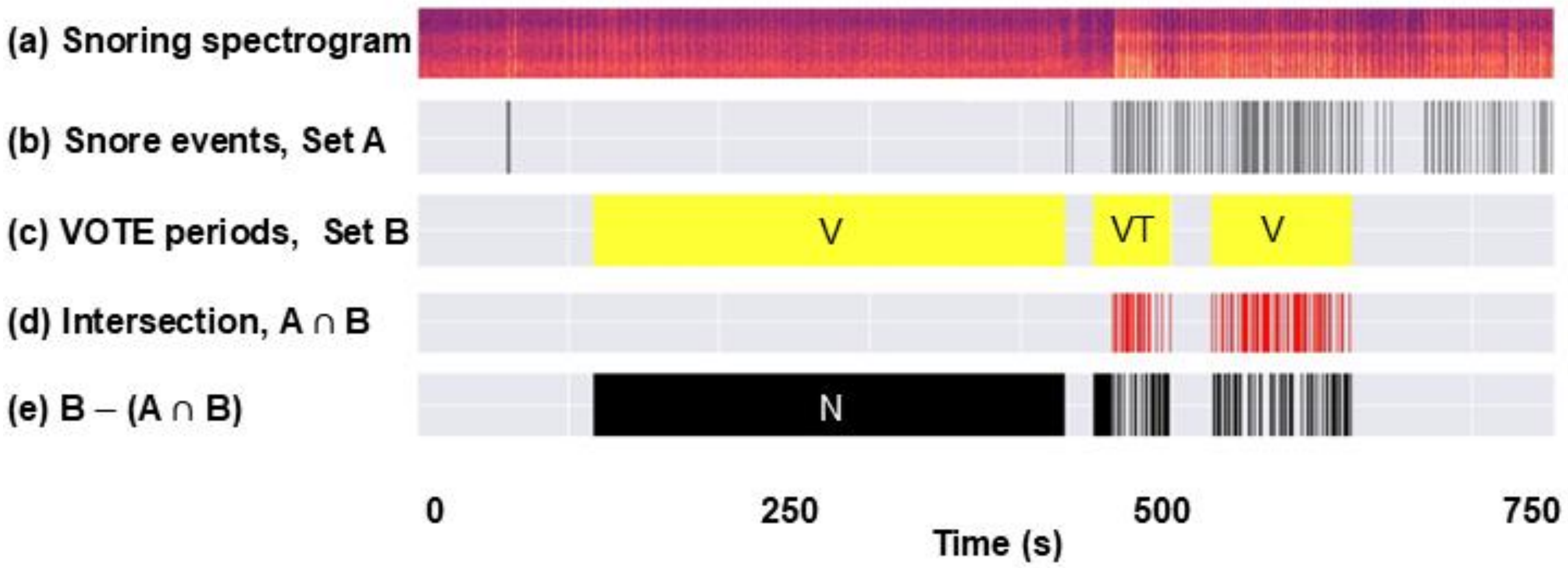

**Figure 2. Illustration of the relationship between the snoring and VOTE labels.**
(a) the spectrogram of a sample snoring sound recording, (b) the labeled snore events, constituting Set A, indicated by dark gray shades, (c) the VOTE-labeled periods, constituting Set B, indicated by yellow shades, (d) the VOTE-labeled snore events formed by the intersection of Set A and Set B, shown as red periods, (e) the non-snoring set obtained by $B - (A \cap B)$, shown as black periods and given a label 'N'.

*3.5 Preparation of Data*

After we had the snore labels (gray periods in Fig 2b; designated as Set A) and VOTE labels (yellow periods in Fig 2c; designated as Set B) aligned in the same timeline, we obtained the intersection of Set A and Set B $(A \cap B)$ to get the VOTE-labeled snoring events (red periods in Fig 2d). The non-snoring periods (black periods in Fig 2e), obtained by $B - (A \cap B)$, were given an "N" label. Note that the first “V”-labeled period (yellow shade labeled as “V” in Fig 2c) did not coincide with any snore labels (gray shades in Fig 2b). It was because a complete obstruction might generate no snoring sounds, or because the slight soft-tissue vibration did not create a sound counted as snoring.

In addition to the VOTE labels, we created another label set depending on the obstruction in the RP and/or RG region: a clip involving any "V" or "O" label was assigned an "RP" label, and a clip associated with any "T" or "E" label was assigned an “RG” label.

Subsequently, we truncated the signals into clips using a 0.5-sec sliding window with a 0.5-sec hop length. If a label (red periods in Fig 2d and black periods in Fig 2e) occupied more than 50% of the duration of a 0.5-sec clip, the clip was kept in the data set. Figure 3 illustrates the process of 0.5-sec clip truncation.

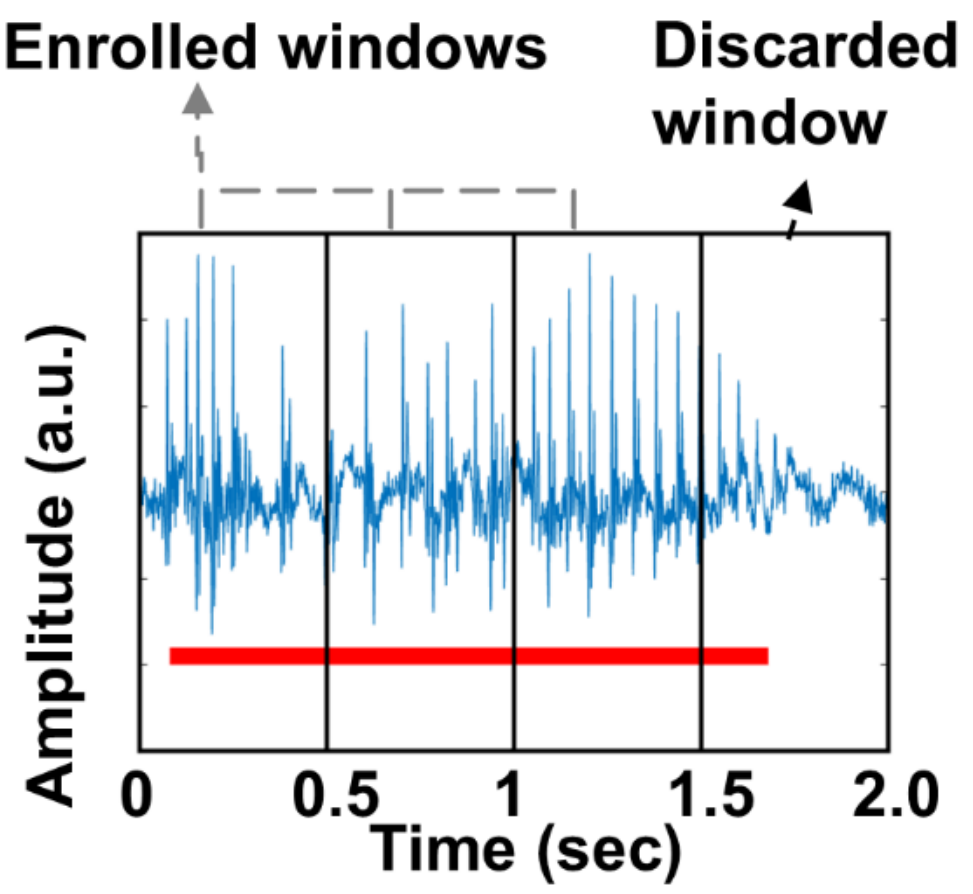

**Figure 3. Illustration of a 0.5-sec clip truncation process.**
The blue trace is a breathing sound signal, and the red line is a label. The sound is divided into 0.5-sec windows, and each 5-sec window is classified as enrolled (the label occupies ≥50% of the 0.5-sec window) or discarded (the label occupies <50% of the 0.5-sec window). For example, the first three of the 5-sec windows are enrolled, but the last is discarded.

*3.6 Multi-label Classification*

In this study, we fine-tuned the pre-trained ResNet-50[28] and AST networks[29] using a multi-label classification approach. While the audio signals of multi-level collapse may exhibit nonlinear coupling and superposition, research in multi-label learning indicates that machine learning/deep learning models can learn label correlations and recover individual labels from compound observation[27, 47]. End-to-end multi-label classification has been shown to reliably recover individual class labels from overlapping acoustic sources without prior source separation, particularly in the context of polyphonic sound event detection[48-50]. ResNet-50 is a classic deep neural network with a residual block that can pass information more efficiently from upstream nodes to downstream nodes in the network. AST adopted an attention mechanism and an encoder-decoder architecture to process acoustic signals using their temporal-spectral representations (spectrograms). Figure 4 shows the architectures of ResNet-50 and AST used in this study. The models' output was a binary classification layer. The binary classification was applied using a threshold δ to determine whether the respective 1) V, O, T, and E obstruction and 2) RP and RG obstruction exist in the input snoring clip.

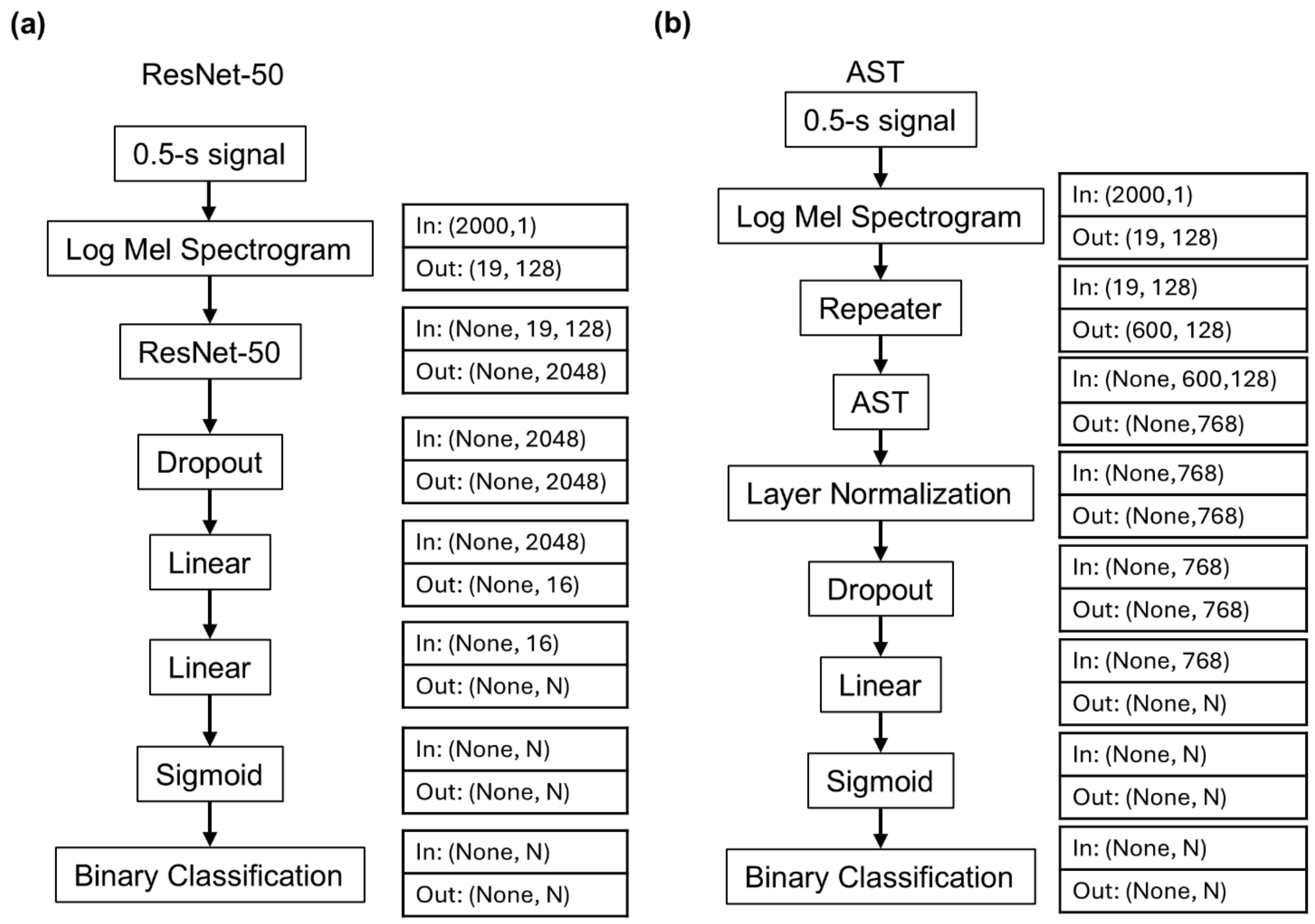

**Figure 4. The architecture of networks used in this study.** (a) ResNet-50 model. (b) AST model. N in this graph stands for the number of classes.

We transformed the 0.5-sec clip into a log mel-spectrogram [51] (window_length=200, hop_length=100, n_fft=1024, n_filters=128, preemp=0.97, window_type='hann', lower_freq=100, and higher_freq=2000). The log mel-spectrogram was further scaled so its mean became 0 and standard deviation became 0.5. The resulting log-mel spectrogram was fed into the model as input.

We adopted a subject-wise assignment strategy to distribute the 0.5-sec clips into the training and test sets to ensure rigorous evaluation and prevent data leakage. The assignment is as random as possible, subject to the condition that the ratio of 0.5-sec clips in the training set to those in the test set was kept as close to 4:1 for each class. SMOTE[30] was employed to generate synthetic samples in the feature space on the training set, a method previously shown to improve classification sensitivity in medical acoustic analysis[52]. Instead of fully balancing the minority classes to match the majority classes, we set a target synthetic capacity of 800 samples for the classes of T and E in the VOTE task and 1400 samples for the class of RG in the RP/RG task, to avoid excessive synthetic noise that could arise from over-interpolating very small clusters. We executed a five-fold cross-validation six times to obtain a more robust report of the model's performance distribution and for statistical comparison.

### *3.7 Training Environment and Hyperparameters*

The training has been done on a GCP virtual machine (OS: SMP Debian 5.10.251-1, CPU: 8 vCPUs Intel Cascade Lake, RAM: 32GB, GPU card: NVIDIA L4). The training process was accelerated using the GPU card, CUDA-12.4, and cuDNN-9.1.

The number of training epochs was 200. The batch size was 16, and the dropout rate was 0.3. The Adam optimizer was adopted. The learning rate was 1e-5, and the momentum was 0.9. Weight decay was 0.0005. An early stop policy was used with a patience of 15 and a plateau patience of 3. To mitigate data imbalance during training, class weights were adjusted based on their ratios. Focal loss cross-entropy was used as the cost function with $\alpha$ set to 0.25 and $\gamma$ set to 2.

*3.8 Evaluation Metrics*

F1-score and Matthew's correlation coefficient (MCC) are crucial metrics used to evaluate the performance of AI models, especially in classification tasks [53, 54]. These metrics are essential for understanding AI model performance, as they provide insights into different aspects of the model's predictive capabilities. MCC considers the true positive (TP) rate, true negative (TN) rate, false positive (FP) rate, and false negative (FN) rate of the classification model, thus more comprehensively reflecting the model's performance, and it is defined as:

$$\text{MCC} = \frac{(TP \times TN - FP \times FN)}{\sqrt{((TP+FP)\times(TP+FN)\times(TN+FP)\times(TN+FN))}} \tag{4}$$

In addition, we used the area under the receiver operating characteristic curve (AUC)[55] to evaluate the models. The holdout test set was used to evaluate the trained models' performance. The models' performance on the test set was reported at the threshold generating the best F1-score during the validation process.

*3.9 Statistical Analysis*

Given that the 30 evaluation measures from 6 runs of 5-fold cross-validation are not independent, we employed the unpaired version of the corrected resampled t-tests[56] to evaluate performance differences among the ResNet-50 and AST models, both with and without SMOTE. This test was employed to adjust variance estimates, thereby reducing the risk of Type I errors (false positives).

Furthermore, to account for the multiple comparison problem, the Bonferroni correction[57] was applied. Effect sizes were measured using Cohen's *d*.

*3.10 Full-Night Monitoring*

To demonstrate the model's potential for monitoring breath sounds during sleep, we recorded the subject's tracheal sounds with the AccurSound AS-101 device over a full night of sleep and performed acoustic analysis retrospectively. The events of inhalation, exhalation, and continuous adventitious sound were identified using a previously proposed convolutional bidirectional gated recurrent network model [58]. The respiratory rate was estimated, and the episodes of apnea (no inhalation for ≥10 seconds) were identified. Assuming that the continuous adventitious sounds were all caused by snoring, the occupation rate [59] of snoring was calculated. Subsequently, the segments of continuous adventitious sound signals were fed into the multi-label VOTE classification model to identify the site of airway obstruction. The results of the analysis were presented using line and bar charts.

## 4. Results

*4.1 Participant Characteristics*

A total of 39 subjects were recruited and enrolled, and two participants' data were excluded from the final analysis, one due to incomplete data and the other because there were no unanimous VOTE labels. The data from 37 subjects were eventually used to fine-tune the models.

The demographics of the 37 subjects (32 males and 5 females) are summarized in Table 2. They are all Taiwanese, with a mean age of 40.05 years and a mean BMI of 27.99. The mean apnea-hypopnea index (AHI) observed was 39.63 events per hour. The severity of obstructive sleep apnea (OSA) varied across the cohort, with one subject exhibiting primary snoring (AHI < 5 events/hour); nine subjects presenting mild OSA (AHI 5–15 events/hour); seven subjects with moderate OSA (AHI 15–30 events/hour); and 20 subjects diagnosed with severe OSA (AHI > 30 events/hour).

**Table 2. Demographic characteristics of the participants.**

| | N = 37 |
|---|---|
| Age (years) | 40.05 |
| Sex | |
| Male | 32 (86%) |
| Female | 5 (14%) |
| Body mass index (kg/m$^2$) | 27.99 |
| Apnea-hypopnea index (events/hour) | 39.63 |
| OSA severity | |
| Normal | 1 (3%) |
| Mild | 9 (24%) |
| Moderate | 7 (19%) |
| Severe | 20 (54%) |

The otolaryngologist labeled nine distinct patterns of airway obstruction based on the VOTE system, ranging from single-level (V, O, T, E) to various multi-level combinations (e.g., VO, VT, OT, OE, and VOT). Detailed distributions of these periods, including total durations and subject counts, are summarized in Table 3. Notably, multi-level obstruction was found in 75.7% of subjects (28/37), and multi-level labels accounted for 59.5% of all labeled periods (3080/5173). The subject-wise distribution of these VOTE-labeled periods is illustrated in Figure 5.

**Table 3. Composition of the VOTE-labeled periods.**

| Obstruction Type | Periods (n) | Duration (s) | Duration (%) | Subjects (n) | Prevalence (%) |
|---|---|---|---|---|---|
| **Single-level** | **35** | **2093** | **40.5** | - | - |
| V | 13 | 1002 | 19.4 | 10 | 27 |
| O | 15 | 793 | 15.3 | 11 | 29.7 |
| T | 4 | 238 | 4.6 | 4 | 10.8 |
| E | 3 | 60 | 1.2 | 2 | 5.4 |
| **Multi-level** | **42** | **3080** | **59.5** | - | - |
| VO | 30 | 2277 | 44 | 21 | 56.8 |
| VT | 2 | 233 | 4.5 | 2 | 5.4 |
| OT | 7 | 387 | 7.5 | 6 | 16.2 |
| OE | 2 | 73 | 1.4 | 2 | 5.4 |
| VOT | 1 | 110 | 2.1 | 1 | 2.7 |
| **Total** | **77** | **5173** | **100** | **37** | - |

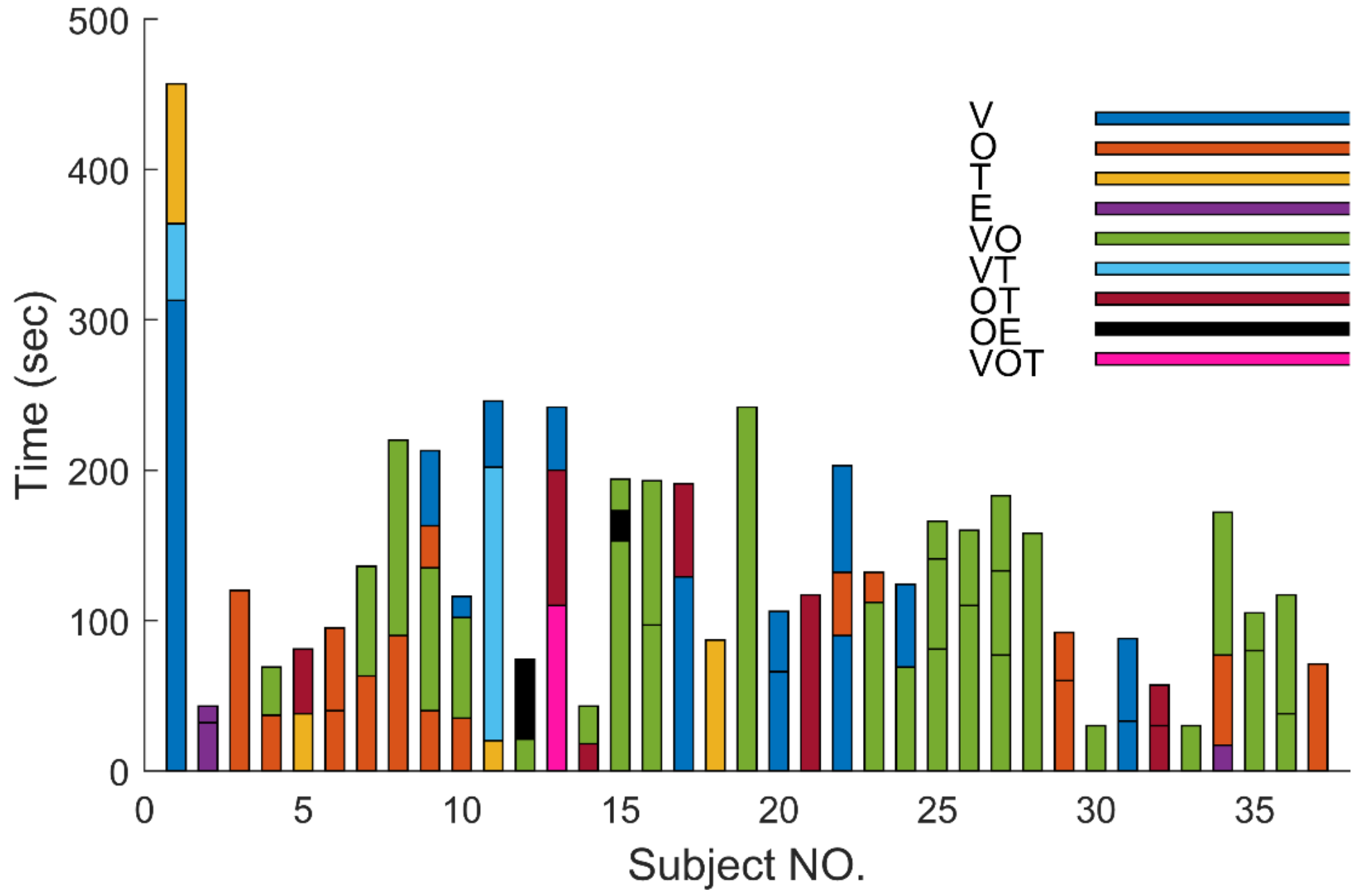

**Figure 5. The distribution of the VOTE-labeled periods among the subjects.**

*4.2 Data Set*

After the signal truncation, our dataset comprised of snoring sound with DISE classification from participants, resulting in 259 V-, 403 O-, 77 T-, 13 E-, 1016 VO-, 46 VT-, 140 OT-, 39 OE-, 30 VOT-, and 3150 N-labeled 0.5-second clips (Table 4). A detailed breakdown of the label distribution within the training and test sets used for the multi-label VOTE and RP & RG classification tasks is also shown in Table 4. Note that the numbers of T, E, and RG labels were disproportionately imbalanced in comparison to V, O, and RP labels.

**Table 4. Summary of the 0.5-sec clips in the training and test sets for the multi-label VOTE and RP & RG classification tasks.**

| Task | Class | Training | Test | Total |
|---|---|---|---|---|
| VOTE | N | 2457 | 693 | 3150 |
| | V | 1137 | 214 | 1351 |
| | O | 1292 | 336 | 1628 |
| | T | 188 | 105 | 293 |
| | E | 23 | 29 | 52 |
| RP & RG | N | 2457 | 693 | 3150 |
| | RP | 1503 | 430 | 1933 |
| | RG | 211 | 134 | 345 |

*4.3 Model Performance*

The performance of ResNet-50 and AST models, trained with/without SMOTE, in identifying snoring sounds associated with obstructions at the V, O, T, and E levels and non-snoring (N) clips is shown in Table 5. The AST model without SMOTE achieved the best performance among the models in identifying snoring sounds at the V level (F1-score: 0.59 ± 0.01, MCC: 0.50 ± 0.02, and AUC: 0.87 ± 0.01) and at the O level (F1-score: 0.76 ± 0.03, MCC: 0.67 ± 0.03, and AUC: 0.92 ± 0.01). The ResNet-50 model with SMOTE achieved the best F1-score and MCC for identifying snoring at the T (0.20 ± 0.09 and 0.14 ± 0.10) and E levels (0.07 ± 0.02 and 0.06 ± 0.05); however, this performance in T and E-level collapse identification was considered very poor. Note that using SMOTE was beneficial for the ResNet-50 model in all V, O, T, and E identification tasks. However, SMOTE did not help the AST model much, and, worse, it confused the model in some cases (e.g., F1-score and MCC for the V- and O-level collapses).

**Table 5. Performance of the ResNet-50 and AST models in VOTE multi-label classification with/without using SMOTE.**

| Metric | Class | Model |
|---|---|---|

| | | Without SMOTE | | | With SMOTE | | |
|---|---|---|---|---|---|---|---|
| | | ResNet-50 | AST | *Cohen's d* | ResNet-50 | AST | *Cohen's d* |
| F1-score | V | 0.41 ± 0.09 | **0.59 ± 0.01**† | 1.83 | 0.45 ± 0.09 | 0.57 ± 0.02 | 1.29 |
| | O | 0.56 ± 0.10 | **0.76 ± 0.03**† | 1.80 | 0.64 ± 0.07 | 0.73 ± 0.03 | 1.06 |
| | T | 0.16 ± 0.06†† | 0.01 ± 0.01 | 2.42 | **0.20 ± 0.09*** | 0.04 ± 0.03 | 1.71 |
| | E | 0.05 ± 0.02 | 0.02 ± 0.03 | 0.69 | **0.07 ± 0.02***** | 0.00 ± 0.00 | 3.52 |
| | N | 0.92 ± 0.01 | **0.93 ± 0.00** | 1.19 | 0.91 ± 0.02 | **0.93 ± 0.00** | 0.82 |
| | Macro Avg | 0.43 ± 0.01 | **0.46 ± 0.00**††† | 2.44 | 0.45 ± 0.00 | **0.46 ± 0.01** | 0.20 |
| MCC | V | 0.21 ± 0.18 | **0.50 ± 0.02**† | 1.53 | 0.30 ± 0.17 | 0.47 ± 0.03 | 0.98 |
| | O | 0.34 ± 0.20 | **0.67 ± 0.03**† | 1.54 | 0.48 ± 0.14 | 0.63 ± 0.04 | 0.91 |
| | T | 0.07 ± 0.08 | 0.00 ± 0.02 | 0.80 | **0.14 ± 0.10** | 0.04 ± 0.03 | 1.03 |
| | E | 0.02 ± 0.03 | -0.01 ± 0.04 | 0.52 | **0.06 ± 0.05** | -0.00 ± 0.00 | 1.20 |
| | N | 0.76 ± 0.02 | 0.79 ± 0.01 | 1.43 | 0.76 ± 0.07 | **0.80 ± 0.01** | 0.69 |
| | Macro Avg | 0.30 ± 0.02 | **0.39 ± 0.00**††† | 4.44 | 0.35 ± 0.01 | **0.39 ± 0.00***** | 3.78 |
| AUC | V | 0.67 ± 0.13 | **0.87 ± 0.01** | 1.50 | 0.71 ± 0.16 | 0.86 ± 0.01 | 0.91 |
| | O | 0.72 ± 0.13 | **0.92 ± 0.01** | 1.50 | 0.81 ± 0.12 | **0.92 ± 0.02** | 0.81 |
| | T | 0.62 ± 0.13 | **0.71 ± 0.06** | 0.72 | 0.70 ± 0.08 | 0.57 ± 0.10 | 1.12 |
| | E | 0.59 ± 0.12 | **0.66 ± 0.10** | 0.44 | 0.63 ± 0.14** | 0.27 ± 0.11 | 2.24 |
| | N | 0.94 ± 0.01 | 0.96 ± 0.00††† | 2.55 | 0.94 ± 0.04 | **0.97 ± 0.00** | 0.78 |
| | Macro Avg | 0.72 ± 0.01 | **0.83 ± 0.00**††† | 10.86 | 0.76 ± 0.01*** | 0.72 ± 0.01 | 3.96 |

Bold numbers indicate the best performance among the four models (ResNet-50 and AST with/without SMOTE). †, †† and ††† stand for a significantly better performance (P <0.05, P<0.01, and P <0.001) between the ResNet-50 and AST models without SMOTE after Bonferroni correction. *, **, and *** stand for a significantly better performance (P <0.05, P <0.01, and P <0.001) between the ResNet-50 and AST models with SMOTE after Bonferroni correction.

Table 6 summarizes the performance of the ResNet-50 and AST models trained with/without SMOTE in identifying snoring sounds at the RP and RG levels. The AST model with/without SMOTE, achieved nearly the same performance in identifying snoring sounds originating from the RP region (F1-score: 0.87, MCC: 0.80, and AUC: 0.97), outperforming the ResNet-50 model. The ResNet-50 model with SMOTE achieved the best F1-score (0.23 ± 0.05) for snoring sound identification at the RG level; however, it did not outperform the AST models in MCC or AUC. Again, SMOTE did not show much help for the AST model. Even though the ResNet-50 model benefited from SMOTE, it remained inferior to the AST model across most performance metrics.

**Table 6. Performance of the ResNet-50 and AST models in RP & RG multi-label classification with/without using SMOTE.**

| Metric | Class | Model | | | | | |
|---|---|---|---|---|---|---|---|
| | | Without SMOTE | | | With SMOTE | | |
| | | ResNet-50 | AST | *Cohen's d* | ResNet-50 | AST | *Cohen's d* |
| F1-score | RP | 0.65 ± 0.05 | **0.87 ± 0.01**††† | 4.22 | 0.79 ± 0.08 | **0.87 ± 0.01** | 0.93 |
| | RG | 0.18 ± 0.04† | 0.09 ± 0.04 | 1.59 | **0.23 ± 0.05***** | 0.08 ± 0.03 | 2.78 |
| | N | 0.79 ± 0.02 | **0.93 ± 0.00**††† | 7.31 | 0.92 ± 0.00 | **0.93 ± 0.01** | 1.17 |
| | Macro Avg | 0.54 ± 0.00 | 0.63 ± 0.00††† | 17.34 | **0.65 ± 0.01***** | 0.62 ± 0.01 | 4.02 |
| MCC | RP | 0.42 ± 0.12 | **0.80 ± 0.01**††† | 3.12 | 0.68 ± 0.18 | **0.80 ± 0.02** | 0.71 |
| | RG | 0.01 ± 0.07 | **0.09 ± 0.05** | 0.85 | 0.07 ± 0.10 | 0.08 ± 0.03 | 0.05 |
| | N | 0.23 ± 0.16 | 0.79 ± 0.01††† | 3.56 | 0.77 ± 0.01 | **0.80 ± 0.02** | 1.42 |
| | Macro Avg | 0.23 ± 0.01 | **0.56 ± 0.01**††† | 28.31 | 0.51 ± 0.01 | **0.56 ± 0.01***** | 2.93 |
| AUC | RP | 0.77 ± 0.10 | **0.97 ± 0.00**†† | 1.99 | 0.89 ± 0.12 | **0.97 ± 0.00** | 0.71 |
| | RG | 0.51 ± 0.10 | 0.75 ± 0.06††† | 2.51 | 0.65 ± 0.06 | **0.78 ± 0.04**** | 1.94 |
| | N | 0.76 ± 0.07 | 0.96 ± 0.00††† | 3.10 | 0.94 ± 0.01 | **0.97 ± 0.00***** | 3.76 |
| | Macro Avg | 0.68 ± 0.01 | 0.89 ± 0.01††† | 21.61 | 0.83 ± 0.01 | **0.91 ± 0.00***** | 9.52 |

Bold numbers indicate the best performance between the models (ResNet-50 and AST with/without SMOTE). †, †† and ††† stand for a significantly better performance (P <0.05, P<0.01, and P <0.001) between the ResNet-50 and AST models without SMOTE after Bonferroni correction. **, and *** stand for a significantly better performance (P <0.01 and P <0.001) between the ResNet-50 and AST models with SMOTE after Bonferroni correction.

*4.4 Full-night Monitoring of Airway Collapse Dynamics and Breath Sounds*

Fig 6 displays the 1st, 2nd, 7th, and 8th hour analysis results of a subject's full-night breath sound monitoring. On the right side of Fig 6a–d, the bar charts show the profile of airway obstruction at the VOTE levels for each hour. We can observe that the model predicted that the snoring sounds mainly originated at the V level, followed by the O level. In the last half hour of sleep, the ratio of the snoring sounds associated with E level rose a bit. Respiratory rate (blue curve), apneic episodes (red marks), and snoring sound occupation rate (purple curve) were drawn on the left side of Fig 6a–d.

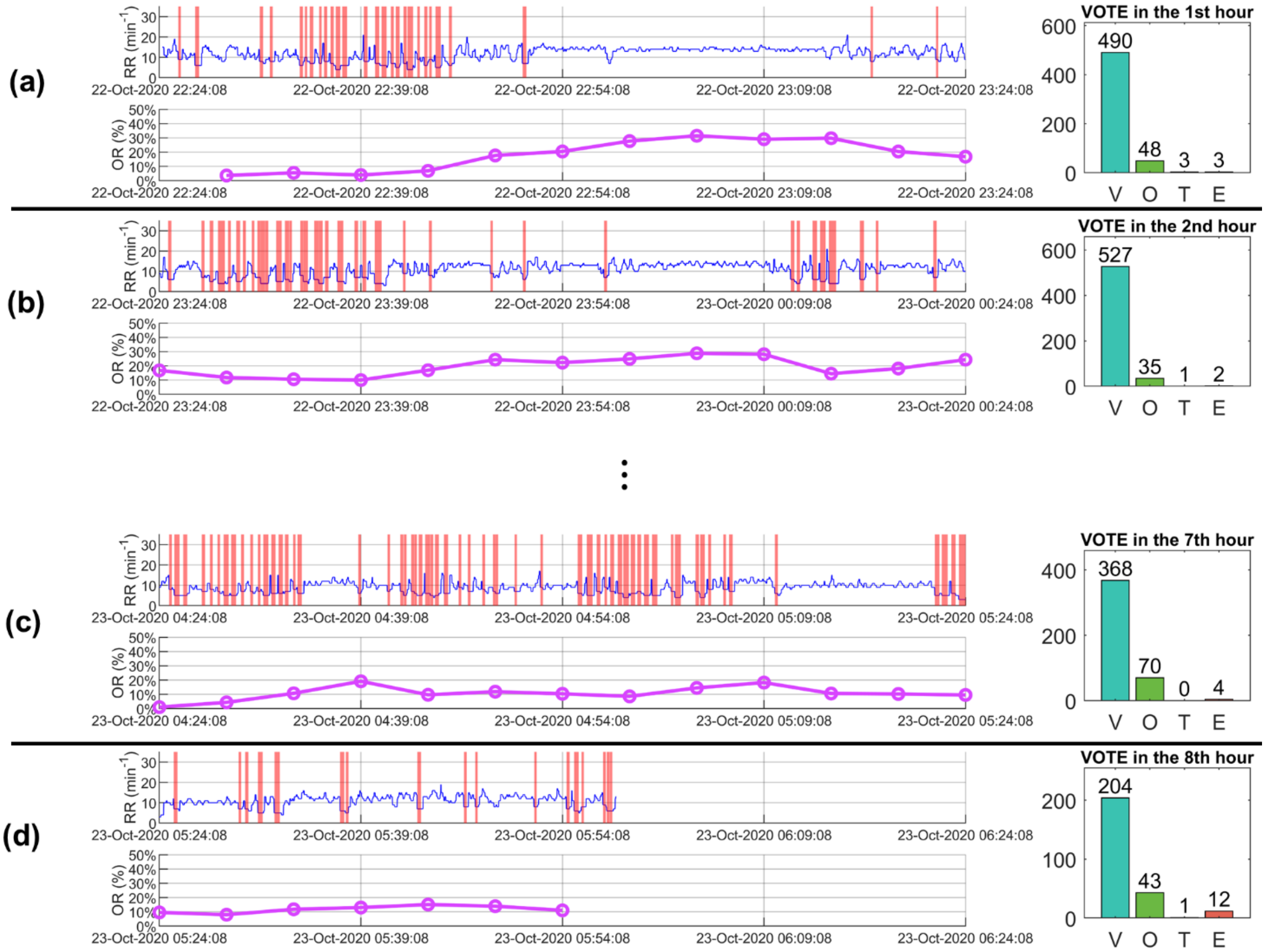

**Figure 6. The (a) first, (b) second, (c) seventh, and (d) 8th-8.5th hour analysis results of a**

**full-night sleep breath sound recording.** The results from the 3rd to 6th hours are not displayed. The upper panel of each subfigure shows the respiratory rate (RR) (blue traces) and apneic episodes (red blocks). The lower panel of each subfigure shows the occupation rate (OR) of snoring sounds (purple line). The right bar charts show the profile of the snoring events originating at the VOTE levels for each hour.

## 5. Discussion

This study fine-tuned the ResNet-50 and AST models in a multi-label classification task to localize snoring sounds in the upper airway in OSA patients. The results showed that multi-level obstruction was observed in 75.7% of subjects and accounted for 59.5% of VOTE-labeled periods, which is consistent with the previous report[23] and highlights the need to transition from traditional single-site analysis to multi-label approaches that capture the complex, concurrent nature of airway collapse. Figure 5 illustrates that a single subject may have both single- and multi-level obstructions present during the DISE procedure.

The AST models demonstrated adequate performance in identifying the snoring sounds associated with the V (F1-score: 0.59 ± 0.01, MCC: 0.50 ± 0.02, AUC: 0.87 ± 0.01) and O levels (F1-score: 0.76 ± 0.03, MCC: 0.67 ± 0.03, AUC: 0.92 ± 0.01) and with the RP level (F1-score: 0.87 ± 0.01, MCC: 0.80 ± 0.01, AUC: 0.97 ± 0.00), as shown in Tables 5 and 6, significantly outperforming the ResNet-50 model. However, the two fine-tuned models failed to identify snoring sounds at the T and E levels, as well as in the RG space. This poor performance is likely due to the limited number of 0.5-sec snoring clips in the T (n = 293) and E (n = 52) classes, as shown in Table 4. Even the use of focal loss, class weight adjustments, and SMOTE can't overcome such a lopsidedly imbalanced and small data set. Nevertheless, the proposed models are still valuable because clinical professionals can easily relate a snoring sound's origin to the T and E levels (or the RG space) if the models indicate no signs of involvement at the V and O levels in the RP space.

As summarized in Table 1, while most previous studies focus on snoring localization at a single level, our approach effectively addresses the more complex challenge of multi-label classification. Our AST model achieved comparable performance in V and O classes—with an F1-score of 0.59–0.87, MCC of 0.50–0.80, and AUC of 0.87–0.97—despite the inherent difficulty of simultaneous multi-site detection compared to the single-level UAR (58.5–87.5%) reported in literature.

Another key distinction of this work is the first-time application of the AST model in snoring localization. Unlike traditional convolutional neural networks (CNNs) or recurrent neural networks (RNNs) architectures that prioritize local patterns, the transformer-based attention mechanism captures long-range spectro-temporal dependencies, which are crucial for distinguishing subtle variations caused by multi-level airway collapse that a local convolutional kernel in CNNs might overlook. Our results demonstrate that AST architecture outperforms the CNN-based ResNet-50, confirming that attention mechanisms offer superior representational power for complex respiratory sound analysis.

Figure 6 illustrates the retrospective breath-sound analysis and the localization of the obstruction site in full-night snoring sounds. The length of the time slice shown in the example is chosen to be one hour. Thus, the profile of airway collapse dynamics is on an hourly basis (the bar charts on the right side). More importantly, the user can more precisely adjust the time slice length and location to match a region of interest if the analysis results are synchronized with other clinical parameters, such as PSG data. By doing so, clinicians can better understand which factors (e.g., sleep stages, body position) induce changes in the dynamics of airway collapse. Even more, such a model can be used in real-time cases. A clinic dentist can immediately determine whether a mandibular advance device effectively alters airway collapse dynamics in a patient lying in a dental chair. Moreover, the models can be deployed on a smartphone as an app. A snorer can easily use a phone to record snoring sounds at home and consult the app for analysis results, facilitating communication with healthcare providers.

This study has limitations. First, the sample size of subjects (N=37) was limited, and data distribution was imbalanced, particularly for the T and E classes, which may have constrained the performance and generalizability of the deep-learning models. Second, the signals were low-pass filtered at 1200 Hz during recording. Even though it may be sufficient to capture the fundamental frequency of high-pitched snoring, the formant information at higher frequencies may be missing. Third, the acoustic features of snoring sounds obtained during DISE may not fully replicate those produced during natural sleep. Induced snoring sounds tend to have a higher frequency component than natural snoring patterns[60]. Propofol sedation also reduces the occurrence of palatal flutter, a significant contributor to snoring sounds [61]. Fourth, the endoscope might interfere with the airway structure and alter the acoustic patterns. Fifth, having a ground-truth VOTE label for each snoring event is difficult. The VOTE period, as labeled by the experts, may contain some noisy labels. Sixth, we did not differentiate between types of obstruction (anteroposterior, laterolateral, or concentric collapses), nor did we consider the severity of the obstruction.

Future research should prioritize expanding the snoring sound data set with VOTE labels and increasing its diversity. The association between snoring acoustics and the types (anteroposterior, laterolateral, and concentric) and extent (partial or complete) of airway obstruction is worth further investigation. The recording device and app should preserve high-frequency spectra, as formants at high frequencies may be important for localizing obstruction sites, such as tongue-base obstruction. The generalizability of the airway collapse monitor system across different patient populations and sleep environments needs further investigation. Additional research is required to integrate and validate the longitudinal snoring tracking system into clinical workflows and to evaluate its impact on patient outcomes. By providing objective, real-time information about airway collapse, clinicians can improve diagnostic accuracy, personalize treatment plans, monitor treatment response, and enhance patient care by enabling more effective, personalized management of the condition.

## Conclusion

This study successfully implemented a multi-label classification framework using the AST to identify multi-level airway obstructions in patients with OSA. Our results demonstrate that the AST model achieves adequate performance in localizing obstructions at the V, O, and RP levels, significantly outperforming ResNet-50. While identifying obstructions at the T and E levels remains challenging due to sample size limitations, this research demonstrates the model's potential for longitudinal tracking of airway collapse dynamics over a full night of sleep, providing a temporal dimension often missing in traditional cross-sectional assessments such as DISE. By bridging advanced deep learning with clinical VOTE classification, this system can aid clinical triage, treatment planning, and longitudinal outcome evaluation. Future work will focus on expanding the dataset diversity to enhance the model's generalizability across all airway levels.

## Acknowledgements

We thank Beluga AI Inc. for technical support and Nian-Jhen Lin, Wan-Ling Tsai, and Yu-Chun You for labeling the snore events.

## Author Contributions

Conceptualization, Y.-C.H., S.Y.-C.L. and F.-S.H.; methodology, J.Y.-J.H., S.-R.H. and Y.-R.C.; software, C.-W.W. and R.-K.C.; investigation, Y.-C.H. and S.Y.-C.L.; resources, Y.-C.H. and F.-S.H.;

data curation, Y.-C.H., S.Y.-C.L. and F.-S.H.; writing—original draft preparation, Y.-C.H. and C.-J.H.; writing—review and editing, S.Y.-C.L. and F.-S.H.; project administration, F.-S.H; funding acquisition, Y.-C.H.

## Statements and Declarations

### *Ethical considerations*

The research was approved by the Research Ethics Review Committee of Taipei Tzu-Chi Hospital (permit number 09-XD-079) on August 3, 2020.

### *Consent to participate*

All participants provided written informed consent before participating in this study.

### *Consent for publication*

All participants agreed that their data could be used for academic publications.

### *Declaration of conflicting interest*

S.-R.H., Y.-R.C., and F.-S.H. are/were employees and own stock in Heroic Faith Medical Science Co., Ltd. The remaining authors have no conflicts of interest to declare.

### *Funding statement*

This study received funding from the Taipei Tzu Chi Hospital, Buddhist Tzu Chi Medical Foundation (Grant No. TCRD-TPE-110-16).

### *Data availability*

Due to pending patent applications and ongoing product development, the raw data that support the findings of this study are not publicly available at this time. Reasonable requests for specific analytical outputs or summary statistics may be considered and should be directed to the corresponding author.